**Electrically tunable total reflection of light by oblique helicoidal cholesteric**

Olena S. Iadlovska[1,2], Kamal Thapa[1,2], Mojtaba Rajabi[1,2], Mateusz Mrukiewicz[1,4], Sergij V. Shiyanovskii[1,3], and Oleg D. Lavrentovich[1,2,3*]

[1]Advanced Materials and Liquid Crystal Institute, Kent State University, Kent, Ohio 44242, USA
[2]Department of Physics, Kent State University, Kent, Ohio 44242, USA
[3]Materials Science Graduate Program, Kent State University, Kent, Ohio 44242, USA
[4]Institute of Applied Physics, Military University of Technology, 00-908 Warsaw, Poland
*Author for correspondence: E-mail: olavrent@kent.edu, tel.: +1-330-672-4844.

**Abstract**

An oblique helicoidal state of a cholesteric liquid crystal (Ch$_{OH}$) is capable of continuous change of the pitch $P$ in response to an applied electric field. Such a structure reflects 50% of the unpolarized light incident along the Ch$_{OH}$ axis in the electrically tunable band determined by $P/2$. Here we demonstrate that at an oblique incidence of light, Ch$_{OH}$ reflects 100% of light of any polarization. This singlet band of total reflection is associated with the full pitch $P$. We also describe the satellite $P/2$, $P/3$, and $P/4$ bands. The $P/2$ and $P/4$ bands are triplets while $P/3$ band is a singlet caused by multiple scatterings at $P$ and $P/2$. A single Ch$_{OH}$ cell acted upon by an electric field tunes all these bands in a very broad spectral range, from ultraviolet to infrared and beyond, thus representing a structural color device with enormous potential for optical and photonic applications.

**Impact statement**

Pigments, inks, and dyes produce colors by partially consuming the energy of light. In contrast, structural colors caused by interference and diffraction of light scattered at submicrometer length scales do not involve energy losses, which explains their widespread in Nature and the interest of researchers to develop mimicking materials. The grand challenge is to produce materials in which the structural colors could be dynamically tuned. Among the oldest known materials producing structural colors are cholesteric liquid crystals. Light causes coloration by selective Bragg reflection at the periodic helicoidal structure formed by cholesteric molecules. The cholesteric pitch and thus the color can be altered by chemical composition or by temperature, but, unfortunately, dynamic tuning by electromagnetic field has been elusive. Here we demonstrate that a cholesteric material in a new oblique helicoidal $Ch_{OH}$ state could produce total reflection of an obliquely incident light of any polarization. The material reflects 100% of light within a band that is continuously tunable by the electric field through the entire visible spectrum while preserving its maximum efficiency. Broad electric tunability of total reflection makes the $Ch_{OH}$ material suitable for applications in energy-saving smart windows, transparent displays, communications, lasers, multispectral imaging, virtual and augmented reality.

**Introduction**

The development of materials capable of dynamically controlled transmission and reflection of light or, more generally, electromagnetic radiation, is one of the most important directions of research with potential applications such as energy-saving smart windows, transparent displays, communications, lasers, multispectral imaging, virtual and augmented reality, etc. [1-4]. A challenging problem is to formulate materials in which light is controlled through the "structural colors" concept, using interference, diffraction, and scattering at submicron-scale architectures, as opposed to absorption. Cholesteric (Ch) liquid crystals (LCs) are among the best-known structurally colored media [5-9]. They are formed by chiral elongated organic molecules that arrange into a helical structure, in which the local molecular orientation $\hat{n}$ twists around a helicoidal axis $\hat{h}$ remaining perpendicular to it. Spatial modulation of the refractive indices with the period $P_0/2$, where the pitch $P_0$ is the distance over which the director $\hat{n}$ rotates by 360 degrees, Fig.1a, enables selective reflection of light [10-12], a property attractive for applications [5-7, 13]. For normal incidence, 50% of unpolarized light is reflected at the half-pitch $P_0/2$; the other 50% are transmitted. The reflected light is circularly polarized with the same handedness as the Ch, while the transmitted light is of opposite circular polarization. Resulting structural Ch colors are highly saturated, they add like colored lights and produce a color gamut greater than that obtained with inks, dyes, and pigments [14]. For oblique incidence of light, more than 50% of light can be reflected, as first reported by Fergason [15] and reviewed recently by Tondiglia et al. [16].

Unfortunately, these structural colors are hard to tune by an electric or magnetic field since the conventional right-angle Ch helix loses its single-mode optical modulation in the presence of the field [9]. One approach is to use polymer-cholesteric composites [13, 16, 17] and to control the structure through the polymer component. For example, one can use a Ch with a positive dielectric anisotropy to switch between a state with selective reflection and a state with polydomain scattering structure [13] or use a Ch with a negative dielectric anisotropy and change the pitch through an electromechanical shift of the ionic polymer network [18, 19].

A direct electromagnetic tuning of Ch structural colors is made possible by the materials with positive dielectric anisotropy that adopt a peculiar oblique helicoidal cholesteric (Ch$_{OH}$) structure [20-29], Fig.1c,d. The Ch$_{OH}$ state forms only in the presence of the electric or magnetic field [20, 21, 30, 31]. While a weak field orients the helicoidal axis $\hat{\mathbf{h}}$ perpendicular to itself forming a light-scattering "fingerprint" Ch state [9], Fig.1b, a field above some threshold $\mathbf{E}_{N^*C}$ realigns $\hat{\mathbf{h}}$ parallel to itself and forms a heliconical Ch$_{OH}$ structure in which $\hat{\mathbf{n}}$ twists around $\hat{\mathbf{h}}$, being also tilted to it [20, 21, 30, 31], Fig.1c,d. The tilt implies that $\hat{\mathbf{n}}$ twists and bends [32]. Because of this, the Ch$_{OH}$ is stable only when the bend elastic constant $K_{33}$ is smaller than its twist $K_{22}$ counterpart, as demonstrated theoretically by Meyer and de Gennes [30, 31]. A crucial advantage of the heliconical Ch$_{OH}$ over the right-angle Ch is that the pitch could be continuously tuned by changing the amplitude of the applied field. Since $\hat{\mathbf{h}}$ is parallel to the field, Ch$_{OH}$ preserves the single-mode character of the refractive index modulation, which guarantees a high efficiency of reflection within a very broad range, from ultraviolet to infrared, including the entire visible spectrum [20-23, 33], Fig.1c,d. At very high fields, the heliconical structure is unwound into a uniformly aligned director [20-22, 33], Fig.1e.

The electrically controlled optical properties of Ch$_{OH}$ have been studied for normal incidence of light [20-23, 33]. At the normal incidence, reflection and transmission are defined by the Ch$_{OH}$ half-pitch $P/2$, similarly to the case of Ch. Oblique propagation of light should reveal the true period $P$ since the propagating wave distinguishes molecular tilts away and towards itself. Oblique incidence of light has been studied theoretically [34-37] and experimentally in the structural analog of Ch$_{OH}$, the chiral smectic C (SmC*) [38-43]. SmC* is formed by molecular layers stacked on top of each other. The molecules are tilted with respect to the normal to layers, so that the director $\hat{\mathbf{n}}$ experiences heliconical modulation along this normal. Berreman [35] simulated numerically propagation of light in SmC* for the incident angle 45° and discovered reflections from *P* and *P*/3, in addition to the reflection from *P*/2 observed also at the normal incidence. The *P*/2 band produces reflections with p- and s-polarizations for both p- and s-polarized incident beams. The reflections at *P* and *P*/3 are different: the p-polarized incident light produces only an s-polarized reflection and vice versa. Berreman admitted that "a convincing intuitive argument for this very odd result" could not be given [34, 35]. Experiments by Hori [38, 39] confirmed the polarization features of reflections at *P* and *P*/2, and demonstrated that the reflection from *P* could reach 100% efficiency for an unpolarized beam [39]. Ouchi et al [36] derived the dispersion relation for the optical eigenmodes in SmC* and demonstrated the observed experimental features of reflections at *P* and *P*/2. The theory was expanded by Oldano et al. [37] who considered the electromagnetic waves as a superposition of eigenmodes, which are the Bloch waves corresponding to the periodic SmC* structure, and predicted that reflections at $P/N$ exhibit different polarization properties for odd and even values of the integer *N*. Namely, even *N*s produce reflection bands in the shape of triplets, whose central peak is common to both eigenmodes and corresponds to total reflection with a polarization exchange; the lateral peaks correspond to the Bragg reflection of each eigenmode. The odd *N*s result in singlets whose characteristics are very similar to the central peak of the triplets [37]. Further experiments by Ozaki et. al. demonstrated the *P*-band reflection tunability by changing the incident angle, the temperature, or the electric field applied perpendicularly to the helicoidal axis $\hat{\mathbf{h}}$ of SmC* [40-43]. The electrically tunable range achieved in Ref. [43] was ~70 nm.

Despite the structural similarities of SmC* and Ch$_{OH}$, their pitch tunability is different. The SmC* exhibits a helicoidal wave of electric polarization; its periodicity could be changed by applying an electric field perpendicularly to $\hat{\mathbf{h}}$. However, the field applied in this direction expands regions with favorable direction of polarization and shrinks the regions of unfavorable polarization, thus destroying the single-mode modulation of the director and diminishing the efficiency of Bragg reflection. This limitation is lifted in the Ch$_{OH}$: since the field acts along the helicoidal axis $\hat{\mathbf{h}}$, it changes the pitch and the conical tilt angle without distorting the single-mode modulation of the optic tensor, which yields highly efficient tunability by hundreds of nanometers [20-22].

In this work, we explore theoretically and experimentally Bragg reflection of light incident obliquely on the Ch$_{OH}$, Fig.1. Previous studies of SmC* are based on numerical modeling. In the present work, taking advantage of a relatively weak birefringence, we develop a coupled wave model with effective resonance matrices, which predicts analytically the locations, intensities, widths, and polarization properties of the reflection bands in the entire range of incident and conical angles. This analytical approach allows one to analyze the properties of reflection as a function of relevant material parameters, Fig.2e-g. This advantage over the purely numerical approach is especially useful for Ch$_{OH}$, in which the material parameters such as the conical angle and pitch can be varied in a broad range by the electric field. The predicted multiple reflection bands corresponding to the pitch $P$, half-pitch $P/2$, one-third of the pitch $P/3$, and a quarter of the pitch $P/4$ are observed and characterized experimentally, Figs.2-6. Bragg reflection at $P$ maintains 100% efficiency for any polarization of light, while its band position is tuned by the electric field in a broad spectral range, embracing hundreds of nanometers, Fig.4a. Since the optic tensor contains only two spatial harmonics corresponding to $P$ and $P/2$, reflections at $P/3$ and $P/4$ are caused by multiple scatterings at $P$ and $P/2$. The observed structure and polarization properties of the Bragg bands are explained qualitatively and quantitatively by the developed model and allows one to determine the pitch and the conical angle of the oblique helicoid. In particular, the difference between the reflection bands with a singlet peak ($N = 1,3$), Figs.2,4, and the triplet peaks ($N = 2,4$), Figs.2,6. is explained by zero diagonal values of the effective resonance matrices for odd $N$s. Broad electric tunability and an efficient 100% reflection make the Ch$_{OH}$ material highly promising for dynamic optical and photonic devices.

**Theory of light propagation in Ch$_{OH}$ structure**

We consider EM wave propagation in a sandwich-type Ch$_{OH}$ cell where the dielectric tensor at optical frequencies (optic tensor) $\boldsymbol{\varepsilon}$ depends only on the $z$ coordinate normal to the substrates, $xz$ is the incidence plane, Fig.1f,g. For a monochromatic wave $\propto e^{-i\omega t}$, the homogeneity in the $xy$ plane preserves the in-plane wave vector $\mathbf{q} = \nu n_g \sin\beta_g \, \hat{\mathbf{x}}$, where $\nu = \omega/c = 2\pi/\lambda$ is the free space wavenumber, $n_g$ is the refractive index of the glass substrate, $\beta_g$ is the incidence angle in the substrate calculated from the $z$-axis, $\hat{\mathbf{x}}$ is the unit vector along the $x$-axis. Using the 4x4 formalism [34], we obtain the solution of the Maxwell equations in the form of 4-vector $z$-dependent amplitudes $(E_x, E_y, H_y, -H_x)\exp(i\mathbf{q}\rho)$, where $\mathbf{F} = (E_x, E_y, H_y, -H_x)$ obeys the equation:

$$\frac{\partial \mathbf{F}}{\partial z} = i\mathbf{L} \cdot \mathbf{F}, \qquad (1)$$

and the 4x4 matrix **L** is

$$\mathbf{L} = \begin{pmatrix} -\frac{q\varepsilon_{31}}{\varepsilon_{33}} & -\frac{q\varepsilon_{32}}{\varepsilon_{33}} & \nu - \frac{q^2}{\nu\varepsilon_{33}} & 0 \\ 0 & 0 & 0 & \nu \\ \nu\left(\frac{\varepsilon_{11}-\varepsilon_{13}\varepsilon_{31}}{\varepsilon_{33}}\right) & \nu\left(\frac{\varepsilon_{12}-\varepsilon_{13}\varepsilon_{32}}{\varepsilon_{33}}\right) & -\frac{q\varepsilon_{13}}{\varepsilon_{33}} & 0 \\ \nu\left(\frac{\varepsilon_{21}-\varepsilon_{23}\varepsilon_{31}}{\varepsilon_{33}}\right) & \nu\left(\frac{\varepsilon_{22}-\varepsilon_{23}\varepsilon_{32}}{\varepsilon_{33}}\right) - \frac{q^2}{\nu} & -\frac{q\varepsilon_{23}}{\varepsilon_{33}} & 0 \end{pmatrix}. \quad (2)$$

In the uniform Ch$_{OH}$, the optic tensor $\boldsymbol{\varepsilon}$ is determined by the constant heliconical angle $\theta$ and the uniformly rotating azimuthal angle $\varphi = \tau z$,

$$\boldsymbol{\varepsilon} = n_o^2 \begin{pmatrix} 1 + 2\delta \sin^2\theta \cos^2\varphi & \delta \sin^2\theta \sin 2\varphi & \delta \sin 2\theta \cos\varphi \\ \delta \sin^2\theta \sin 2\varphi & 1 + 2\delta \sin^2\theta \sin^2\varphi & \delta \sin 2\theta \sin\varphi \\ \delta \sin 2\theta \cos\varphi & \delta \sin 2\theta \sin\varphi & 1 + 2\delta \cos^2\theta \end{pmatrix}. \quad (3)$$

where $\tau = \frac{2\pi}{P}$ is the main harmonic of the periodic structure and $\delta = \frac{(n_e^2 - n_o^2)}{2n_o^2} \approx \frac{(n_e - n_o)}{n_o}$ is the relative birefringence, $n_e$ and $n_o$ are the extraordinary and ordinary refractive indices, respectively. When $\theta = const$, $\varepsilon_{33} = const$ and Equation (1) can be reduced to a dimensionless wave equation for $(E_x, E_y)$ by introducing the dimensionless coordinate $\tilde{z} = \tau z$ and expressing $H_y$ and $H_x$ from the first and second rows of Equation (1), respectively:

$$E_x'' + 2ib_{1x}(2\cos\tilde{z} E_x' + \sin\tilde{z} E_y') + (a_x^2 - 2ib_{1x}\sin\tilde{z} + 2b_{2x}\cos 2\tilde{z})E_x +$$
$$2(ib_{1x}\cos\tilde{z} + b_{2x}\sin 2\tilde{z})E_y = 0, \quad (4)$$
$$E_y'' + 2ib_{1y}\sin\tilde{z} E_x' + 2b_{2y}\sin 2\tilde{z} E_x + (a_y^2 - 2b_{2y}\cos 2\tilde{z})E_y = 0,$$

where $' = \partial/\partial \tilde{z}$ and a factor 2 simplifies the equations for Fourier harmonics below. The coefficients in Equation (4) are defined by $\delta$, $\theta$, the normalized wavenumber $\eta = \frac{n_o \nu}{\tau} = \frac{n_o P}{\lambda}$, and the 'ordinary' angle of light incidence $\beta_o = \arcsin\left(\frac{q}{n_o \nu}\right) = \arcsin\left(\frac{n_g \sin\beta_g}{n_o}\right)$ which is very close to the angle $\beta_g$ of light propagation in the glass; furthermore,

$$a_x^2 = \eta^2 g_x (\cos^2\beta_o (1 + \delta \sin^2\theta) + 2\delta \cos^2\theta),$$
$$a_y^2 = \eta^2 g_y \cos^2\beta_o (\cos^2\beta_o + \delta + \delta \cos^2\theta), \quad (5)$$
$$b_{1x} = b_1 g_x, \quad b_{2x} = b_2 g_x, \quad b_{1y} = b_1 g_y, \quad b_{2y} = b_2 g_y,$$

where $g_x^{-1} = 1 + 2\delta \cos^2 \theta$, $g_y^{-1} = \cos^2 \beta_o + 2\delta \cos^2 \theta$, $b_1 = \frac{\delta \eta}{2} \sin \beta_o \sin 2\theta$ and $b_2 = \frac{\delta \eta^2}{2} \cos^2 \beta_o \sin^2 \theta$.

Equation (4) is a system of linear ordinary differential equations with periodic coefficients. Therefore, its solution is a Bloch wave expressed as a series of harmonics:

$$\mathbf{E}(\tilde{z}) = \begin{pmatrix} E_x(\tilde{z}) \\ E_y(\tilde{z}) \end{pmatrix} = e^{ik\tilde{z}} \sum_{m=-\infty}^{\infty} \mathbf{E}^{(m)} e^{im\tilde{z}}, \tag{6}$$

where the vector amplitudes $\mathbf{E}^{(m)}$ obey the equation

$$\sum_{p=-2}^{2} \mathbf{M}_p(m) \mathbf{E}^{(m+p)} = 0, \tag{7}$$

with the matrix $\mathbf{M}_p(m)$ representing the $p^{\text{th}}$ harmonic of the periodic optical properties of Ch$_{\text{OH}}$,

$$\mathbf{M}_0(m) = \begin{pmatrix} a_x^2 - (k+m)^2 & 0 \\ 0 & a_y^2 - (k+m)^2 \end{pmatrix},$$

$$\mathbf{M}_{\pm 1}(m) = \begin{pmatrix} -b_{1x}(2k + 2m \mp 1) & \mp ib_{1x}(k+m) \\ -ib_{1y}(1 \pm (k+m)) & 0 \end{pmatrix}, \tag{8}$$

$$\mathbf{M}_{\pm 2}(m) = \mathbf{M}_{\pm 2} = \begin{pmatrix} b_{2x} & \pm ib_{2x} \\ \pm ib_{2y} & -b_{2y} \end{pmatrix}.$$

The four non-trivial solutions of Equation (7) are invariant under the transformation $\mathbf{E}^{(m)} \to \mathbf{E}^{(m+l)}$ and $k \to k - l$, thus we analyze the solutions in which $\mathbf{E}^{(0)}$ have the *largest* amplitudes.

Since $b_{n\xi} \propto \delta \approx 0.1$ (here, the subscripts adopt all possible values, $n = 1, 2$, $\xi = x, y$), $\mathbf{M}_p(m)$ with $p \neq 0$ are small and the harmonics $\mathbf{E}^{(m)}$ are mostly controlled by $u = (a_x^2 + a_y^2)/2 = n_{eff}^2 \eta^2$ and $v = (a_x^2 - a_y^2)/2 = \delta_{eff} \eta^2$, where $n_{eff}$ and $\delta_{eff}$ are the homogeneous effective refractive index and birefringence, respectively,

$$n_{eff}^2 = \cos^2 \beta_o + \frac{\delta}{2} \left( \frac{\sin^2 \theta \cos^2 \beta_o}{(\cos^2 \beta_o + 2\delta \cos^2 \theta)} + \frac{4 \cos^4 \theta - \cos^2 \beta_o (3 \cos 2\theta + 1)}{2(1 + 2\delta \cos^2 \theta)} \right),$$

$$\delta_{eff} = \frac{\delta \sin^2 \beta_o [\cos^2 \beta_o (3 \cos 2\theta + 1) + 4\delta \cos^4 \theta]}{4(1 + 2\delta \cos^2 \theta)(\cos^2 \beta_o + 2\delta \cos^2 \theta)} \approx \frac{1}{4} \delta \eta^2 \sin^2 \beta_o (3 \cos 2\theta + 1). \tag{9}$$

Note that $\delta_{eff}$ and $v$ (i) change the sign at $\theta \approx 53°$ [44], being positive in the Ch$_{\text{OH}}$ with $\theta < 33°$ and negative in the right-angle Ch with $\theta = 90°$, (ii) decrease when $\beta_o$ decreases and completely vanish for the normal incidence.

**Far from the $N^{\text{th}}$ order Bragg** conditions $k = N/2$, one can reduce Equation (7) to

$$\bar{\mathbf{M}}_0 \mathbf{E}^{(0)} = 0 \tag{10}$$

for the dominant harmonic $\mathbf{E}^{(0)}$, where $\bar{\mathbf{M}}_0 = -k^2\mathbf{I} + \check{\mathbf{M}}$, $\mathbf{I} = \text{diag}(1,1)$ is the 2d identity matrix, and $\check{\mathbf{M}}$ is the perturbation expansion in $\delta$ that results in the polynomial:

$$\check{\mathbf{M}} = \text{diag}(a_x^2, a_y^2) - \sum_{l \neq 0} \mathbf{M}_l(0)\mathbf{M}_0^{-1}(l)\mathbf{M}_{-l}(l) + O(\delta^3), \tag{11}$$

The eigenmodes' wavevectors $k_j$ are the eigenvalues for $\check{\mathbf{M}}$,

$$k_{\mp,\pm} = \mp\sqrt{\frac{1}{2}\left(\check{M}_{1,1} + \check{M}_{2,2} \pm \sqrt{(\check{M}_{1,1} - \check{M}_{2,2})^2 + 4\check{M}_{1,2}\check{M}_{2,1}}\right)}, \tag{12}$$

and determine their polarizations through Equation (10); here and below, subscripts $\mp$ and $\pm$ on the left-hand side of Equation (12) correspond to $\mp$ and $\pm$ on the right-hand side, respectively.

If $\beta_o > \beta_c$, where $\sin^2\beta_c \approx \left|\frac{\delta\eta^3 \sin^4\theta}{(1-\eta^2)(3\cos 2\theta+1)}\right|$ is determined from the condition $(\check{M}_{1,1} - \check{M}_{2,2})^2 = 4\check{M}_{1,2}\check{M}_{2,1}$, then the propagating forward eigenmodes are linearly polarized, being the ordinary s mode $E_{x,z}^{(0)} = 0$ with $k_s = k_{+,-} \approx \sqrt{\check{M}_{2,2}} \approx a_y$ and the extraordinary p mode $E_y^{(0)} = 0$ with $k_p = k_{+,+} \approx \sqrt{\check{M}_{1,1}} \approx a_x$. When $\beta_o < \beta_c$ and decreases, the ellipticity increases, reaching the circular polarization at normal incidence, $\beta_o = 0°$, where $\check{M}_{1,1} = \check{M}_{2,2}$.

**Bragg reflections.** The $N^{\text{th}}$ order Bragg reflection is caused by the resonant increase of $\mathbf{E}^{(-N)}$ up to the amplitude of the major harmonic $\mathbf{E}^{(0)}$ when $a_x, a_y \approx \frac{N}{2}$ and $k \approx \frac{N}{2}$. To model this process, we derive the equations for $\mathbf{E}^{(0)}$ and $\mathbf{E}^{(-N)}$ by eliminating non-resonance harmonics $\mathbf{E}^{(l)}$, $l \neq 0, -N$, from Equation (7):

$$\widetilde{\mathbf{M}}_0(0)\mathbf{E}^{(0)} + \widetilde{\mathbf{M}}_{-N}(0)\mathbf{E}^{(-N)} = 0,$$
$$\widetilde{\mathbf{M}}_N(-N)\mathbf{E}^{(0)} + \widetilde{\mathbf{M}}_0(-N)\mathbf{E}^{(-N)} = 0, \tag{13}$$

where the resonance matrices $\widetilde{\mathbf{M}}_j(m)$ are the polynomial expansions in $\delta$,

$$\widetilde{\mathbf{M}}_j(m) = \mathbf{M}_j(m) - \sum_{l \neq j,0} \mathbf{M}_l(m)\mathbf{M}_0^{-1}(l+m)\mathbf{M}_{j-l}(l+m) + O(\delta^3). \tag{14}$$

The first term on the right-hand side of the Equation (14) is dominant in the first, $N = 1$, and the second, $N = 2$, order Bragg reflections, and it is absent in all higher order reflections, $N \geq 3$, Equation (8). These higher order reflections are caused by the indirect interaction between the resonant harmonics through the non-resonant ones and are described in subsequent terms on the right-hand side of the Equation (14). To find the behavior of the four eigenmodes near the $N^{\text{th}}$ order, the determinant of the Equation (13) is evaluated into the fourth order polynomial expansion in a small parameter $\mu_N = k - \frac{N}{2}$,

$$\mu_N^4 - 2B_N \mu_N^2 + C_N = 0, \tag{15}$$

where $B_N$ and $C_N$ are functions of $u_N = u - \left(\frac{N}{2}\right)^2$, which change the sign in the $N^{\text{th}}$ order Bragg reflection region, whereas $v \propto \delta \eta^2$, Equation (9), and $b_{n\xi} \propto \delta \eta^n$, $n = 1,2$, Equation (5). The solutions of Equation (15) produce two evanescent modes when $C_N < 0$ and all four eigenmodes when (i) $D_N = B_N^2 - C_N < 0$ or (ii) $B_N < 0$ and $C_N > 0$. The results for the first four Bragg reflection orders are presented below.

**First order, $N = 1$.** For the first order Bragg diffraction, Equation (15) has the coefficients

$$B_1 = u_1^2 + v^2 - \frac{\bar{b}_1^2}{4}, \quad C_1 = \left(u_1^2 - v^2 - \frac{\bar{b}_1^2}{4}\right)^2 \geq 0, \tag{16}$$

where

$$\bar{b}_1 = \sqrt{b_{1x} b_{1y}} = \frac{\delta \eta \sin \beta_o \sin 2\theta}{2\sqrt{(1+2\delta \cos^2 \theta)(\cos^2 \beta_o + 2\delta \cos^2 \theta)}} \tag{17}$$

and results in four eigenmodes with the wavevectors

$$k_{\mp,\pm} = \frac{1}{2} \mp v \pm \sqrt{4u_1^2 - \bar{b}_1^2}. \tag{18}$$

All four eigenmodes are evanescent with the same $\alpha = \left|\text{Im } k_{\mp,\pm}\right|$ in the region $u_1^r < u_1 < u_1^b$, where $u_1^{b,r} = \pm \bar{b}_1/2$ are the blue and red edges of the Bragg reflection band, respectively. Thus, $\bar{b}_1$ determines both the width of the first order Bragg reflection band $\Delta u_1 = u_1^b - u_1^r = \bar{b}_1$ and its maximum $\alpha_{max} = \bar{b}_1$ at the center, $u_1 = 0$. Since $\bar{b}_1 \propto \tan \beta_o \sin 2\theta$, the first order Bragg reflection band (i) becomes narrower and weaker when $\beta_o$ decreases and completely disappears for the normal incidence, (ii) reaches maximum when $\theta = 45°$ and disappears in the right-angle Ch, $\theta = 90°$, and in the unwound helicoid, $\theta = 0°$.

The incident from $z < 0$ light generates the mode with $k_{-,+}$ and harmonics $E_y^{(0)}$ and $E_{x,z}^{(-1)}$, and mode with $k_{+,+}$ and harmonics $E_{x,z}^{(0)}$ and $E_y^{(-1)}$. Because the evanescent modes contain harmonics with crossed polarizations and have the same $\alpha = \left|\text{Im } k_{\mp,\pm}\right|$, the incident s-polarized wave reflects as the p-polarized one, and vice versa, while the full reflectivity remains polarization independent.

To compare the model results with the experimental spectra presented below, the plots of multiple order Bragg bands in Ch$_{OH}$ and Ch, Figs.2e-g and Fig.6d, are shown in scaled coordinates $\hat{u}_N = u_N/\delta\eta^2$ and $\hat{\alpha} = \alpha/\delta\eta^2$ to make the plots more general because $v \propto \delta\eta^2$ and $b_2 \propto \delta\eta^2$ for the second order; for the first order, $\bar{b}_1 \propto \delta\eta$ requires an additional factor $\eta = \sqrt{N^2+4u_N}/2n_{eff}$ and $\bar{b}_1/\delta\eta^2$ has no singularity when $\beta_o \to 90°$.

**Second order, $N = 2$.** For the second order Bragg reflection, the coefficients in biquadratic Equation (15) are

$$B_2 = \frac{2(u_2^2+v^2)-b_{2+}^2}{8}, C_2 = \frac{(u_2^2-v^2)^2-(b_{2+}u_2+b_{2-}v)^2}{16}, \quad (19)$$

where $b_{2\pm} = (b_{2y} \pm b_{2x})/2$. The eigen wavevectors

$$k_{\mp,\pm} = 1 \pm \sqrt{B_2 \mp \sqrt{D_2}} \quad (20)$$

are evaluated using the standard analysis of a biquadratic equation; here

$$D_2 = B_2^2 - C_2 = [(u_2 v + b_{2+}b_{2-})^2 - (v^2 - b_{2+}^2)(b_{2+}^2 - b_{2-}^2)]/4 \quad (21)$$

is the discriminant of Equation (15).

The properties of eigenmodes are primarily controlled by the sign of $C_2$. Equation $C_2 = 0$ has four real solutions $u_{2-}^- < u_{2-}^+ < u_{2+}^- < u_{2+}^+$,

$$u_{2\mp}^\pm = \mp\sqrt{(v \pm b_{2-})^2 + b_{2x}b_{2y}} \pm b_{2+} \quad (22)$$

presented by magenta, red, blue, and black lines, respectively, in half-pitch Bragg bands of Ch$_{OH}$ and Ch, Figs.2f,g; thus, the entire range of $u_2$ splits into five regions with alternating signs of $C_2$. In the outer regions $(u_2 < u_{2-}^-)$ and $(u_2 > u_{2+}^+)$ where $C_2 > 0$, all wavevectors $k_{\mp,\pm}$, Equation (20), are real and represent the propagating modes; thus, Bragg reflection does not occur.

In the magenta-red $(u_{2-}^- < u_2 < u_{2-}^+)$ and blue-black $(u_{2+}^- < u_2 < u_{2+}^+)$ side regions, a negative $C_2$ results in a negative $B_2 - \sqrt{D_2} = -\alpha_-^2$ and a complex $k_{-,\pm} = 1 \pm i\alpha_-$, Equation (20), while $k_{+,\pm}$ remain real. Thus, incident light exhibits the single mode Bragg reflection bands in both regions, with the maxima at

$$u_{2\mp}^m = \frac{1}{2}(u_{2\mp}^+ + u_{2\mp}^-). \quad (23)$$

The width of the reflection bands is the same as the width of the corresponding regions $\Delta u_{2\mp} = u_{2\mp}^+ - u_{2\mp}^-$, top and bottom parts of the Fig.2f, respectively. Both bands are caused by the same mode with $k_{-,+} = 1 + i\alpha_-$. Under the condition $|v| > b_{2+}$ that is obeyed in Ch$_{OH}$ when $\beta_o > \theta$ and in the right-angle Ch when $\beta_o > 60°$, this mode has linearly p-polarized harmonics $E_{x,z}^{(0)}$ and $E_{x,z}^{(-1)}$ in magenta-red region for Ch$_{OH}$, Fig. 2f, and in blue-black region for Ch, Fig.2g, respectively, whereas this mode has s-polarized harmonics $E_y^{(0)}$ and $E_y^{(-1)}$ in the alternative region. When $\beta_o$ decreases, the harmonics' polarization in both modes transforms into an elliptical one with the same sign and further into the same circular polarization for the normal incidence, $\beta_o = 0°$, while the bands coalesce into the single band of the exact solution in both conventional right-handed Ch and Ch$_{OH}$.

The central red-blue region ($u_{2-}^+ < u_2 < u_{2+}^-$) contains four areas. The area encircled by the green line corresponds to $D_2 < 0$ and determines the band similar to the first order Bragg reflection band where all four eigenmodes have the same $\alpha = |\text{Im } k_{\mp,\pm}|$ and consist of the cross-polarized harmonics. The central band's maximum position

$$u_{2c}^m = -\frac{b_{2+}b_{2-}}{v} \approx -\frac{\delta\eta^2(1+\cos^2\beta_o)\sin^4\theta}{4(1+3\cos2\theta)} \tag{24}$$

and its width

$$\Delta u_{2c} = \sqrt{b_{2x}b_{2y}(1-v^{-2}b_{2+}^2)} \approx \frac{1}{2}\delta\eta^2 \cos\beta_o \sin^2\theta \sqrt{1 - \frac{\sin^4\theta(2-\sin^2\beta_o)}{\sin^4\beta_o(1+3\cos2\theta)^2}} \tag{25}$$

are defined by the roots of $D_2 = 0$, Equation (21), and explain their dependencies on $\beta_o$ for Ch$_{OH}$, Fig.2f, and Ch, Fig.2g. Similar to the first order Bragg reflection band, the incident s-polarized wave in this band reflects as the p-polarized one, and vice versa.

In the area enclosed by red, blue and green lines, all four eigenmodes are also evanescent because $C_2 > 0$ and $B_2 < 0$, $B_2 \mp \sqrt{D_2} = -\alpha_\mp^2$ and $k_{\mp,\pm} = 1 \pm i\,\alpha_\mp$ (20), Fig.2f and Fig.2g.

In the two remaining areas enclosed by red-green and blue-green lines in Fig.2f and Fig.2g, respectively, all wavevectors $k_{\mp,\pm}$, Equation (20), are real and represent the propagating modes.

**Third order, $N = 3$.** Bragg reflection of the third order occurs when $a_{x,y} \approx \frac{3}{2}$. The matrices

$$\widetilde{\mathbf{M}}_{-3}(0) = \widetilde{\mathbf{M}}_3(-3) = \begin{pmatrix} \frac{1}{4}b_{1x}(7b_{2y} - 12b_{2x})\mu_3 & -i\frac{3}{4}b_{1x}b_{2-} \\ i\frac{3}{4}b_{1y}b_{2-} & \frac{5}{4}b_{1x}b_{2y}\mu_3 \end{pmatrix} \tag{26}$$

control the interaction between the resonant harmonics $\mathbf{E}^{(0)}$ and $\mathbf{E}^{(-3)}$, Equation (13), and are similar to the matrices $\widetilde{\mathbf{M}}_{-1}(0)$ and $\widetilde{\mathbf{M}}_1(-1)$ in the first order Bragg reflection. $\widetilde{\mathbf{M}}_{-3}(0)$ and $\widetilde{\mathbf{M}}_3(-3)$ also have zero-value diagonal elements when $\mu_3 = 0$ and generate the coefficients in Equation (15) that obey the condition $C_3 \geq 0$,

$$B_3 = \frac{\tilde{u}_3^2 + v_3^2 - \bar{b}_3^2}{9}, \quad C_3 = \frac{(\tilde{u}_3^2 - v_3^2 - \bar{b}_3^2)^2}{81} \geq 0, \tag{27}$$

where $\tilde{u}_3 = u_3 - \delta u_3$, $v_3^2 = \left(v - b_{1x}b_{1-} + \frac{b_{2-}b_{2+}}{5}\right)^2 + g_x g_y \left(b_{1x}b_1 + \frac{b_{2-}b_2}{5}\right)^2$, $\bar{b}_3^2 = \frac{9 b_{1x} b_{1y} b_{2-}^2}{16}$, and $\delta u_3 = \frac{b_{2+}^2}{5} - b_{1x}^2 \frac{-7 b_{1x} b_{1y}}{16}$. The four eigenmodes near the third order Bragg reflection have the wavevectors

$$k_{\pm,\pm} = \frac{3}{2} \pm \frac{v_3}{3} \pm \frac{1}{3}\sqrt{\tilde{u}_3^2 - \bar{b}_3^2} \tag{28}$$

and are all evanescent when $\tilde{u}_3^2 < \bar{b}_3^2$. Similar to the first-order Bragg reflection, the evanescent modes contain harmonics with crossed polarizations; therefore, the incident p mode reflects as the s-mode and vice versa.

Parameter $\delta u_3 = \frac{b_{2+}^2}{5} - b_{1x}^2 \frac{-7 b_{1x} b_{1y}}{16}$ determines the center of the Bragg peak, whereas

$$\bar{b}_3 = \frac{3\sqrt{b_{1x} b_{1y}} b_{2-}}{4} = \frac{\delta^2 \eta^3 \sin^3 \beta_o \cos^2 \beta_o \sin^2 \theta \sin 2\theta}{4\sqrt{(1+2\delta \cos^2 \theta)^3 (\cos^2 \beta_o + 2\delta \cos^2 \theta)^3}} \tag{29}$$

defines its maximum value $\mathrm{Im}\, k_{\pm,+} (u_3 = \Delta u_3) = \frac{\bar{b}_3}{3}$ and its width $\Delta u_3 = u_3^b - u_3^r = 2\bar{b}_3$, where $u_3^b = \delta u_3 + \bar{b}_3$ and $u_3^r = \delta u_3 - \bar{b}_3$ are the blue and the red edges of the band, respectively. Equation (29) for $\bar{b}_3$ demonstrates the absence of the third order Bragg reflection for the normal incidence, consistent with the exact solution. It suggests that the experimental observation of the third order Bragg reflection in Ch$_{\mathrm{OH}}$ is facilitated by a weak field (which implies a larger $\theta$) and by a large incidence angle $\beta$, Fig.2d.

**Fourth order, $N = 4$.** The Bragg reflection of the fourth order occurs when $a_{x,y} \approx 2$ and is very similar to the second order: the matrices $\widetilde{\mathbf{M}}_{-4}(0)$ and $\widetilde{\mathbf{M}}_4(-4)$ control the interaction between the resonant harmonics $\mathbf{E}^{(0)}$ and $\mathbf{E}^{(-4)}$, Equation (13), and are proportional to the second order matrices $\mathbf{M}_{\pm 2}$,

$$\widetilde{\mathbf{M}}_{-4}(0) = g_4 \mathbf{M}_{-2}, \quad \widetilde{\mathbf{M}}_4(-4) = g_4 \mathbf{M}_{+2}, \tag{30}$$

with the coefficient $g_4 = \frac{\delta \sin^2 \beta_o \sin^2 \theta (\cos^2 \beta_o + 2\delta \cos^2 \theta)}{2[(1+\delta \sin^2 \theta) \cos^2 \beta_o + 2\delta \cos^2 \theta][\cos^2 \beta_o + \delta(1 + \cos^2 \theta)]}$. Thus, the coefficients $B_4$ and $C_4$ in biquadratic Equation (15) are proportional to the second order coefficients $B_2$ and $C_2$,

$$B_4 = \frac{2(u_4^2 + v^2) - b_{4+}^2}{8}, \quad C_4 = \frac{(u_4^2 - v^2)^2 - (b_{4+} u_4 + b_{4-} v)^2}{16}, \tag{31}$$

where $b_{4\pm} = g_4 b_{2\pm}$, and the eigenmodes for the fourth order Bragg reflection have the same three-band structure with polarization characteristics described by Equations (19) – (25), in which the index 2 is substituted with 4. The strong dependence of the $g_4$ factor on the conical angle $\theta$ and the angle of incidence $\beta_o$ implies that the observation of the fourth order Bragg reflection in Ch$_{OH}$ with a small $\theta$ is difficult. As will be shown in the experimental part, Fig.6e, this peak is visible only when the field is weak (so that $\theta$ is not very small) and $\beta_o$ is large. The $g_4$ factor is responsible for the difference in the $\beta_o$ dependence of the fourth and second order reflections in both Ch$_{OH}$ and Ch. In Fig.2g, which describes the second order Bragg reflection of the conventional Ch at $P_0/2$, the decrease of $\beta_o$ to zero produces a single mode band, which corresponds to the exact solution. In contrast, the decrease of $\beta_o$ to zero in the case of the fourth order $P_0/4$ reflection of Ch results in a disappearance of all three bands, see Fig.6d.

One would expect that all orders of Bragg reflection should have three bands corresponding to the resonant interaction of p-p, s-p (p-s), and s-s modes. However, for odd $N$, the matrices $\widetilde{M}_{-N}(0)$ and $\widetilde{M}_N(-N)$ in Equation (13) have zero-value diagonal elements when $\text{Re } k = \frac{N}{2}$, providing the condition $C_N \geq 0$ and making p-p and s-s bands in the odd order Bragg reflections non-existent.

**Experimental Results**

**Multiple bands of reflection.** According to theory, Bragg reflection of the $N$th order at the wavelength $\lambda_{P/N} \approx \frac{n_o P \cos\beta_o}{N}$ appears whenever $a_x, a_y \approx \frac{N}{2}$, where the angle $\beta_o$ of light propagation in the liquid crystal is calculated using Snell's law, $\beta_o = \arcsin\left(\frac{n_g \sin\beta}{n_o}\right)$, and the angle $\beta$ of light incidence is measured from the normal to the Ch$_{OH}$ cell, Fig.1f,g. The reflection spectra are measured for different values of $\beta$ and the electric field $E$; the field tunes the heliconical angle $\theta$ and the pitch $P$, Fig. 1c,d, and shifts reflection bands corresponding to the full pitch $P$ and its fractions $P/2$, $P/3$, and $P/4$. The reflections at $P$, $P/2$, and $P/3$ are measured in the Ch$_{OH}$ dopped with 4 wt.% of the chiral additive (abbreviated as M4), and reflection at $P/4$ is characterized using the Ch$_{OH}$ with 3 wt.% of the chiral dopant (M3).

Experimental spectrum in Fig.2a shows three orders of Bragg reflection, $N = 1, 2,$ and $3$, of an unpolarized light that impinges obliquely, $\beta = 55°$, at the Ch$_{OH}$ stabilized by the field, $E =$ 0.7 V/μm. In addition to the conventional reflection at $P/2$, one observes a reflection band at $P$ and a peculiar reflection at $P/3$. The $P$, $P/2$, and $P/3$ bands are well separated from each other. The $P$-band centered at $\lambda_P$ = 930 nm, and the $P/3$ band at $\lambda_{P/3}$ = 360 nm both represent a single maximum, while the reflection at $P/2$ exhibits three peaks. The ratio $\lambda_P/\lambda_{P/3}$ = 2.58 is different from 3 because of the dispersion of refractive indices, Fig.S1. The reflection at $P/4$ is discussed later since this band is in the UV region for a given electric field and cannot be shown in Fig. 2.

**Polarization properties**. Electric control of the pitch allows one to shift the reflection bands and explore their polarization properties in a convenient part of the spectrum, Fig.2b-d. In agreement with the theory, the reflection band at $P/2$ is a triplet, Equations (23) and (24) and Fig.2f, while reflections at $P$, Equation (18) and Fig.2e, and $P/3$, Equation (29), are single peaks. When the incident light is p-polarized, its reflection at $P$ and $P/3$ is s-polarized, and vice versa, an s-polarized incident ray changes into a p-polarized one, Fig.2b,d. The $P/2$ band is more complicated, with three maxima, Fig. 2c, as expected from Fig.2f. The central peak exhibits a conversion of the p (s) polarization of the incident light into an s (p) polarization of the reflected light, similar to the singlet peaks of reflection at $P$ and $P/3$. The blue-shifted peak of the $P/2$ triplet shows a strong reflection of the s-polarized light into a similarly s-polarized light, while the red-shifted maximum corresponds to the reflection of p-polarized light into a similarly p-polarized beam, Fig.2c.

All bands in Fig.2b-d show secondary reflections, such as finite s-s reflection outside the Bragg bands, and satellite peaks in the second-order Bragg reflection in Fig.2c. The weak s-s reflection in Fig.2b-d is caused by reflection at the electrodes at both glass plates. The p-p reflection is much weaker at large incidence angles as it is close to the Brewster condition. As a result, the s-s reflection is always larger than the p-p reflection outside the strong reflection band p-s in Fig.2b and 2d. Inside the p-s band, the s-s and p-p reflections are weaker than the ones outside the band since the contribution from the opposite substrate is eliminated.

Figure 2c shows the satellite s-s and p-p peaks located under the main p-s peak. This result is unusual since these peaks cannot be caused by an ideal Ch$_{OH}$ structure. We attribute the effect to the fact that the subsurface structure (the conical angle and the pitch) of the Ch$_{OH}$ is different from an ideal helicoid because of the surface anchoring [24]. As a result, the incident light becomes elliptically polarized, which means that the p-component of the incident light reflects as the s-components and vice versa. This effect brings about the additional peaks such as the s-s and p-p reflections underneath of the p-s band in Fig.2c. A similar effect is the appearance of a weak p-s and s-p reflections under the s-s peak and even weaker reflections under the p-p band in Fig.2c.

**Heliconical angle $\theta$ and pitch $P$ of the Ch$_{OH}$.** The structure of the Bragg reflection at $P/2$ allows one to determine the heliconical angle $\theta$ and the pitch $P$ directly from spectra, such as one in Fig.2c. The peak wavelengths $\lambda_{P/2}$ with p-p, p-s, and s-s polarization states, Fig.3a, are substituted into Equations (23) and (24) to calculate $\theta$ and $P$, Figs. 3b and 3c, respectively. The peak wavelength $\lambda_{P/2}$ at a given polarization is determined by measuring the peak's bandwidth at its half-amplitude and defining the middle coordinate of the bandwidth. We use notations with superscripts to identify the state of polarization of the incident and reflected beams; for example, $\lambda_{P/2}^{ps}$ refers to a p-polarized incident light that is reflected as an s-polarized beam. The wavelength $\lambda_{P/2}^{ps} = \lambda_{P/2}^{sp}$ corresponds to the central peak in the triplet, Equation (24). The angle $\theta$ and the pitch $P$ are determined from the equations

$$\lambda_{P/2}^{\sigma} = n_{o\sigma} P \sqrt{U_{\sigma} - \Delta U_{\sigma}}, \tag{32}$$

by solving for the pairs $\{\lambda_{P/2}^{pp}, \lambda_{P/2}^{ss}\}$, $\{\lambda_{P/2}^{pp}, \lambda_{P/2}^{ps}\}$, and $\{\lambda_{P/2}^{ps}, \lambda_{P/2}^{ss}\}$, where $\sigma = $ pp, ps, ss,

$$U_{\sigma} = \frac{(\cos^2 \beta_o (1+\delta_{\sigma} \sin^2 \theta) + 2\delta_{\sigma} \cos^2 \theta)}{2(1+2\delta_{\sigma} \cos^2 \theta)} + \frac{\cos^2 \beta_o (\cos^2 \beta_o + \delta_{\sigma} + \delta_{\sigma} \cos^2 \theta)}{2(\cos^2 \beta_o + 2\delta_{\sigma} \cos^2 \theta)},$$

$$\Delta U_{ps} = -\frac{\delta_{ps}(1+\cos^2 \beta_o) \sin^4 \theta}{(16-24 \sin^2 \theta)},$$

and for p-p and s-s polarizations,

$$\Delta U_{\sigma} = \pm \frac{\delta_{\sigma}}{8} \left( \sqrt{8 \sin^4 \beta_o \cos^2 \theta (2 - 3 \sin^2 \theta) + (1 + \cos^2 \beta_o)^2 \sin^4 \theta} + \sqrt{8 \sin^4 \beta_o \cos 2\theta (2 - 3 \sin^2 \theta) + (1 + \cos^2 \beta_o)^2 \sin^4 \theta} \right),$$

with + and – corresponding to pp and ss peak wavelengths, respectively. Here $\delta_{\sigma} = \frac{(n_{e\sigma}^2 - n_{o\sigma}^2)}{2n_{o\sigma}^2}$, where refractive indices $n_{o\sigma} = n_o(\lambda_{P/2}^{\sigma})$ and $n_{e\sigma} = n_e(\lambda_{P/2}^{\sigma})$ are chosen for the corresponding peak wavelength. The values of $\theta$ determined from (32) for pairs $\{\lambda_{P/2}^{pp}, \lambda_{P/2}^{ss}\}$, $\{\lambda_{P/2}^{pp}, \lambda_{P/2}^{ps}\}$ and $\{\lambda_{P/2}^{ps}, \lambda_{P/2}^{ss}\}$ are in a good agreement with each other, Fig.3b. This result is remarkable since $\theta$ is determined by the difference in the two peaks' positions and is thus sensitive to how accurately these are measured, Fig.2c. The pairs $\{\lambda_{P/2}^{pp}, \lambda_{P/2}^{ss}\}$ with the largest separation and $\{\lambda_{P/2}^{pp}, \lambda_{P/2}^{ps}\}$, where the peaks are the narrowest and of the most symmetric shape, provide a more reliable $\theta$ than $\{\lambda^{ps}_{P/2}, \lambda^{ss}_{P/2}\}$, in which case the wide asymmetric shape of $\lambda^{ss}_{P/2}$ is the probable source of deviations seen in Fig.3b. The determination of pitch $P$ is robust, yielding practically the same values when measured from all three pairs $\{\lambda^{pp}_{P/2}, \lambda^{ss}_{P/2}\}$, $\{\lambda^{pp}_{P/2}, \lambda^{ps}_{P/2}\}$, and $\{\lambda^{ps}_{P/2}, \lambda^{ss}_{P/2}\}$, Fig.3c.

Figure 3d shows how the change of the electric field changes the number of the cholesteric pseudolayers. The parameter $\widetilde{\Delta}_{i+1} = \widetilde{N}(E_i) - \widetilde{N}(E_{i+1})$ represents the difference between the number of pseudolayers defined experimentally as $\widetilde{N} = d/P$ at two close values of the applied electric field $E_i$ and $E_{i+1}$. This parameter should be an integer if the change of the structure occurs by an insertion of a full period of the Ch$_{OH}$ structure. However, the experiment shows that this parameter is not always exactly integer. The reason is in the complex $z$-dependence of the structure, with the ideal heliconical deformation distorted near the substrates, as explained previously [24]. The field dependencies of $P$ and $\theta$ in Fig.3 are obtained from Bragg reflection and therefore represent the values $P_{bulk}$ and $\theta_{bulk}$ in the homogeneous bulk region of the cell; these parameters are different near the substrates. The true change of the number of pseudolayers is shown in Fig.3d as a parameter $\Delta_{i+1}$ defined as the integer closest to $\widetilde{\Delta}_{i+1}$. When $E < 0.9$ V/μm, the two parameters practically coincide, $0 < \widetilde{\Delta} - \Delta \ll 1$ and the number of pseudolayers changes by 1; for higher fields, when the pitch is small, the number of added pseudolayers is often higher, $\Delta = 2,3$, and the difference $\widetilde{\Delta} - \Delta$ becomes substantial. The reason for this high-field behavior is that the decrease of $P$ amplifies the effect of the inhomogeneous subsurface layers.

The formulae (32) represent an independent and direct approach to measure $\theta$ and $P$ from spectra. A different approach, based on the elastic and dielectric properties of the Ch$_{OH}$, was proposed earlier [20], in which $P$ and $\theta$ are theoretically predicted (assuming they do not change along the normal to the cell) as [20]:

$$P = \frac{\kappa E_{NC} P_0}{E} = \frac{2\pi}{E}\sqrt{\frac{K_{33}}{\varepsilon_0 \Delta \varepsilon}}; \qquad (33)$$

$$\sin^2 \theta = \frac{\kappa}{1-\kappa}\left(\frac{E_{NC}}{E} - 1\right), \qquad (34)$$

where $\kappa = K_{33}/K_{22}$, $E_{NC} = 2\pi K_{22}/P_0\sqrt{\varepsilon_0 \Delta \varepsilon K_{33}}$ is a critical field at which the Ch$_{OH}$ transforms into an unwound state with $\theta = 0$, Fig.1e. Therefore, one can combine the spectral, Equation (32), and the elasto-dielectric, Equations (33) and (34), approaches for the same cell and determine, in addition to $\theta$ and $P$, also other parameters such as $\kappa$ and $\frac{K_{33}}{\Delta\varepsilon}$.

**Electric field tunability**. The unique feature of the Ch$_{OH}$ P-band is that it produces a total reflection of obliquely incident light and that the wavelength $\lambda_P$ of this reflection is tunable in a very broad range, for example, from 400 nm to 700 nm in Fig.4a, without diminishing its 100% efficiency. Both $\lambda_P$ and the bandwidth $\Delta\lambda_P$ decrease at higher fields, Fig.4b, since P and θ both decrease, as expected from Equations (33)-(34). The electric field dependencies in Fig.4b are measured for a p-polarized incident light; the reflected beam is s-polarized. The data are in a good agreement with the theory; peak wavelengths at the full pitch P (N = 1) satisfy the condition $(a_x^2 + a_y^2) = \frac{1}{2}$, Equation (15), where $a_x$ and $a_y$ are analytical expressions in Equation (5) with material parameters $P_0$, $E_{NC}$, and $\kappa$, measured previously [33], Fig.4b. The reflection band at P/3 shows a similar tunability by the electric field, Fig.4c. To observe the P/3 band in the visible range, the electric field is lowered, which increases the pitch. The electric field dependency of the spectral position of the P/3 band is similar to that of the P-band; the peak wavelength $\lambda_{P/3}$ and the bandwidth $\Delta\lambda_{P/3}$ both decrease with the field, Fig.4d. Similarly to $\lambda_P$, the $\lambda_{P/3}$ data are in a good agreement with the theory (N = 3), Fig.4d. Note that the dependencies in Fig.4c have been measured in thick cells with a hybrid alignment in order to enhance the intensity of P/3 band, Equation (29), and to change the pitch continuously.

**Dependence on the incident angle**. Figure 5 shows that the wavelength of the Bragg peak for P and P/3 bands could be controlled by the angle of light incidence β. A higher β causes blue shifts of the reflection wavelength since $\lambda_P, \lambda_{P/3} \propto \cos\beta$, Fig.5a. For a fixed field $\lambda_P, \lambda_{P/3} \propto \cos\beta$. At a constant field, $E = 1.1$ V/m, when β increases from 30° to 58°, the total reflection peak changes from about 610 nm to 450 nm, while its bandwidth increases, Fig.5a,b. The P/3 reflection band shows a similar blue shift as β increases, Fig.5c. The experimental data are well fitted with the theoretical model with the parameters indicated in the Fig.5 caption.

**Even order Bragg diffraction peaks**. Triplet reflections at the half of Ch$_{OH}$ pitch (P/2), Fig.2c, and at the quarter of Ch pitch ($P_0/4$), Fig.6a, are similar. The central peak in both the Ch and Ch$_{OH}$ triplets corresponds to the reflection that alters the polarization of light: if the incident light is p-polarized, the reflected beam is s-polarized, and vice versa. The lateral peaks preserve the polarization of the incident beams. An important difference is that in Ch$_{OH}$, the lateral peak corresponding to the p-p reflection is located on the red side of the band, Fig.2c, while in the Ch case, the p-p peak is blue-shifted, Fig.6a. The s-s peaks show a complementary behavior, being blue-shifted in Ch$_{OH}$ and red-shifted in Ch. The effect is caused by the fact that spatially-averaged Ch$_{OH}$ is a positive uniaxial birefringent material whereas Ch is a negative one, with the optic axis in both cases being along the twist axis, Equation (9), Fig.2f and Fig.6d.

Observation of the $P/4$ band in Ch$_{OH}$ is more difficult than in Ch since its intensity is very low, $\alpha \propto g_4 b_2 \propto sin^4\theta$, Equations (5) and (30). The $P/4$ band is observed in a planar cell, $d = 16.2$ μm, filled with M3 (which yields a higher pitch) at $\beta = 52.5°$, Fig.6e. The reflection at $P/4$ is caused by double scattering at the $P/2$. The peak wavelengths $\lambda_{P/4}^{ps}$ is in a good agreement with the calculated 4$^{th}$ Bragg order, in which the central peak of $P/4$ satisfies the condition $(a_x^2 + a_y^2) = 8$ when $N = 4$, Equations (5) and (15). Due to low reflection intensity, polarization properties of the $P/4$ band are difficult to analyze.

**Conclusions**

We demonstrate that a single thin sandwich-type cell filled with an oblique helicoidal Ch$_{OH}$ and acted upon by an electric field enables total reflection of obliquely incident light, which is tunable in a very broad spectral range, from ultraviolet to infrared. In a particular example in Fig.4a, the tunable range is from 400 to 700 nm. As demonstrated recently, a single cell filled with the Ch$_{OH}$ materials could yield electric tunability of the reflected structural color from some short wavelength $\lambda$ all the way up to the longest wavelength ~20 $\lambda$ [33]. The tunable total reflection of obliquely incident light is associated with the period $P$ of Ch$_{OH}$. The electric field applied parallel to the heliconical axis of Ch$_{OH}$ tunes the value of $P$ but does not change the single-mode character of the refractive index modulation, which explains why the efficiency of reflection maintains its 100% value in a broad spectral range, Fig.4a. This 100% reflection is observed for any polarization of incident light.

In addition to the 100% reflection at $P$, the Ch$_{OH}$ shows multiple orders of Bragg reflection, corresponding to $P/2$, $P/3$, and $P/4$. The $P/3$ and $P/4$ reflections are associated with the multiple scattering events at $P$ and $P/2$. Both theoretical model and experiment establish polarization patterns of the reflections, which are triplets in the case of the $P/2$ and $P/4$ bands and singlets in the $P$ and $P/3$ cases. The model allows one to extract the material parameters such as the heliconical angle $\theta$ and the pitch $P$ directly from the triplet structure of the $P/2$ band, Fig.3a-c. The knowledge of these two parameters allows one to determine the elastic constants of twist $K_{22}$ and bend $K_{33}$ using Equations (33) and (34) and optical spectra, without the need to resort to the measurements of the unwinding electric field $E_{NC}$.

Total reflection occurs when all four eigenmodes are evanescent and the wavevectors are complex. The most favorable conditions for the total reflection are created in the first order Bragg reflection band that exists for oblique light incidence in the Ch$_{OH}$, Fig.2b,e, but does not exist in the right-angle Ch. This band exists for any incidence angle except for normal incidence ($\beta_o = 0$). It exhibits features useful for applications. Namely, the polarization of reflected light is orthogonal to the polarization of the incident light. Such a feature might be useful to eliminate noise reflection which typically preserves the state of polarization. The coefficient of reflection does not depend on the polarization of the incident light and remains the same for linearly, circularly, or unpolarized incident light. The strong reflectivity is achieved for thin, 20-40 µm, samples thanks to the large effective amplitude $\bar{b}_1$ of optical tensor modulation.

The demonstrated total reflection of obliquely incident light of an arbitrary state of polarization and its broad tunability by an electric field suggest that Ch$_{OH}$ materials will find applications in dynamic tunable optical and photonic devices. Among many advantages that they offers is the simplicity of construction and operation: a relatively thin (20-40 µm) layer of a Ch$_{OH}$ sandwiched between two plates with planar transparent electrodes is already a device capable of extraordinary broad multispectral tunability.

**Materials and Methods**

*Ch$_{OH}$ materials.* A room-temperature nematic (N) LC precursor is formulated using flexible dimers 1″,7″-bis(4-cyanobiphenyl-4′-yl) heptane (CB7CB), and 1-(4-cyanobiphenyl-4′-yloxy)-6-(4-cyanobiphenyl-4′-yl) hexane (CB6OCB) (both purchased from SYNTHON Chemicals GmbH & Co. KG), and a rod-like mesogen pentyl cyanobiphenyl (5CB) (EM Industries), in the weight proportion 5CB:CB7CB:CB6OCB = 52:31:17. Upon cooling, the N mixture shows the phase diagram N$_{TB}$ (17.4 °C) N (78 °C) I, where N$_{TB}$ is the twist-bend nematic [45-49]. The dimer molecules yield a small $K_{33}$ [45-49], while 5CB shifts the temperature range of the mesophase down to room temperature. The Ch$_{OH}$ mixtures are prepared by doping N with the left-handed chiral agent S811 (EM Industries), 5CB:CB7CB:CB6OCB:S811 = 49:31:17:3 (abbreviated as M3) and 5CB:CB7CB:CB6OCB:S811 = 50:30:16:4 (abbreviated as M4). The phase diagrams of M3 and M4 are as follows: N$_{TB}^*$ (17.1 °C) Ch (65.0 °C) I, and N$_{TB}^*$ (15.5 °C) Ch (66.3 °C) I, respectively. Here, $N_{TB}^*$ is the chiral analog of the twist-bend nematic [45-49]. The phase diagram is measured on cooling using Linkam hot stage equipped with the PE94 temperature controller (both Linkam Scientific) and EHEIM Professional-3 cooling system (EHEIM GmbH & Co. KG). The spectra at $Ch_{OH}$ are measured without the hot plate at room temperatures, 19 °C – 20.5 °C, with the accuracy ± 0.2 °C.

*Cells with preset surface alignment.* Two types of flat sandwich cells are studied, with the planar and hybrid [26] anchoring of the local director at the bounding surfaces. In planar cells, the $Ch_{OH}$ structure adjusts its pitch slightly in order to comply with the boundary conditions. In hybrid aligned cells, one substrate yields a planar alignment while the other produces a homeotropic alignment. The homeotropic substrate helps to tune the structure in a continuous manner since it imposes no in-plane axis for the director $\hat{\mathbf{n}}$. Planar cells with indium tin oxide (ITO) electrodes and gap thicknesses $d = (16 – 52)$ μm were either purchased from E.H.C. Co, Japan, or assembled in the laboratory. Planar alignment is achieved by a rubbed layer of polyimide PI2555 (Nissan Chemicals, Ltd.). Homeotropic substrates are prepared in the laboratory using the technique developed by Young-Ki Kim at Kent State [50, 51]. The reactive mesogen RM257 is mixed with the homeotropic aligning agent SE5661 and doped with the UV photo-initiator Irgacure 651, in weight proportion SE5661:RM257:Irgacure 651 = 96.7:3:0.3. The mixture is spin-coated onto an ITO glass substrate and baked at 170 °C for 1 hour. The substrates are then irradiated with UV light (365 nm) for 90 minutes [50, 51]. The cell thickness is set by spherical spacers mixed with UV-curable adhesive NOA68 (Norland Products, Inc.) and measured using Lambda 18 UV/VIS spectrometer (Perkin Elmer, Inc). The cells are filled with the $Ch_{OH}$ mixture in the isotropic phase and slowly cooled down at the rate 0.2°C/min to the room temperature.

*Electric field control of the structure.* To create the $Ch_{OH}$ state, an alternating current (ac) field with a frequency 3 kHz is applied to the ITO electrodes at the bounding plates using Keithley 3390 waveform function generator (Keithley Instruments) and wideband Krohn-Hite 7602M amplifier (Krohn-Hite Co.), Fig.1g. The field **E** establishes the $Ch_{OH}$ state with a helicoidal axis $\hat{\mathbf{h}}$ parallel to itself, $\hat{\mathbf{h}}||\mathbf{E}$, Fig.1c,d; the $Ch_{OH}$ pitch $P$ changes in a broad range in response to the changing amplitude of **E** [20, 21].

*Spectral measurements at an oblique incidence of light.* The scheme of the experimental setup is shown in Fig.1f. The fiber optics setup is based on a rotating goniometer stage (Euromex Holland), and optical spectra are measured using a tungsten halogen light source LS-1 with a working range 350-2000 nm and a USB2000 VIS-NIR spectrometer (both Ocean Optics, Inc.). Polarization of the incident and reflected (or transmitted) beams is set by a pair of linear wire-grid polarizers (Thorlabs, Inc.) placed along the optical path at the exit from the fiber connected to the light source and at the entrance to the fiber connected to the USB2000 detector. The incident ray is p-polarized when its electric field $\mathbf{E}_p$ oscillates in the plane of incidence $xz$ and s-polarized when the electric field $\mathbf{E}_s$ is perpendicular to it. The Ch$_{OH}$ cell is mounted in the center of the goniometer stage, and the angle of light incidence $\beta$ is measured from the normal $\hat{\mathbf{z}}$ to the Ch$_{OH}$ cell with an accuracy better than 0.25°. The Ch$_{OH}$ cell is sandwiched between two semi-cylindrical prisms (N-BK7 glass, $n_p = 1.518$ at 547 nm, Thorlabs, Inc.). A Cargille matching liquid with the refractive index $n = 1.516$ (at 600 nm) fills the gaps between the cell substrates and prisms to reduce the reflection losses. The semi-cylindrical prisms allow one to measure light reflection in a broad angular range and assure that the angle of light incidence from the glass is the same as the incidence angle from the air: $\beta_g = \beta_{air} \equiv \beta$. Optical spectra are normalized with respect to the intensity of light $I_0$ passed through the Ch$_{OH}$ cell in the transmission mode when the structure is unwound by a high applied field $E = 7$ V/μm; this value exceeds the critical field $E_{NC}$ of unwinding the Ch$_{OH}$ into the N state with the director $\hat{\mathbf{n}}$ along the field, Fig.1e. $E_{NC}$ is measured by the capacitance method described previously [33]. Reflection spectra are measured at $E < E_{NC}$, when the material adopts the heliconical Ch$_{OH}$ state, Fig.1c,d. The reflection coefficient is calculated as $R[\%] = \frac{I_R}{I_0} \times 100\%$, where $I_R$ is the intensity of light reflected from the Ch$_{OH}$.

**Funding**

The work was supported by the NSF grant ECCS-1906104 and ECCS-2122399 (theory and experiments) and by a grant from Oculus/Meta (materials optimization). MM's visit to Kent State was supported by the Polish Ministry of Science and Higher Education under the Mobilność Plus program (grant no. 1644/MOB/V/2017/0) and the Kosciuszko Foundation exchange program.

**Author contributions**

OI measured spectra at $Ch_{OH}$ and analyzed the data; KT prepared cells and measured the refractive indices of N; MR assisted in the measurement of birefringence and analyzed the data; MM measured preliminary data for Fig.2b,c using a similar $Ch_{OH}$ composition; SVS conceived the project and developed a theory; ODL conceived and directed the project; OI, SVS and ODL wrote the manuscript with the inputs from all co-authors. All authors contributed to scientific discussions.

**Competing interests:** The authors declare no competing interests.

**Availability of data and material:** All data are included in the main text and the electronic supplementary material.

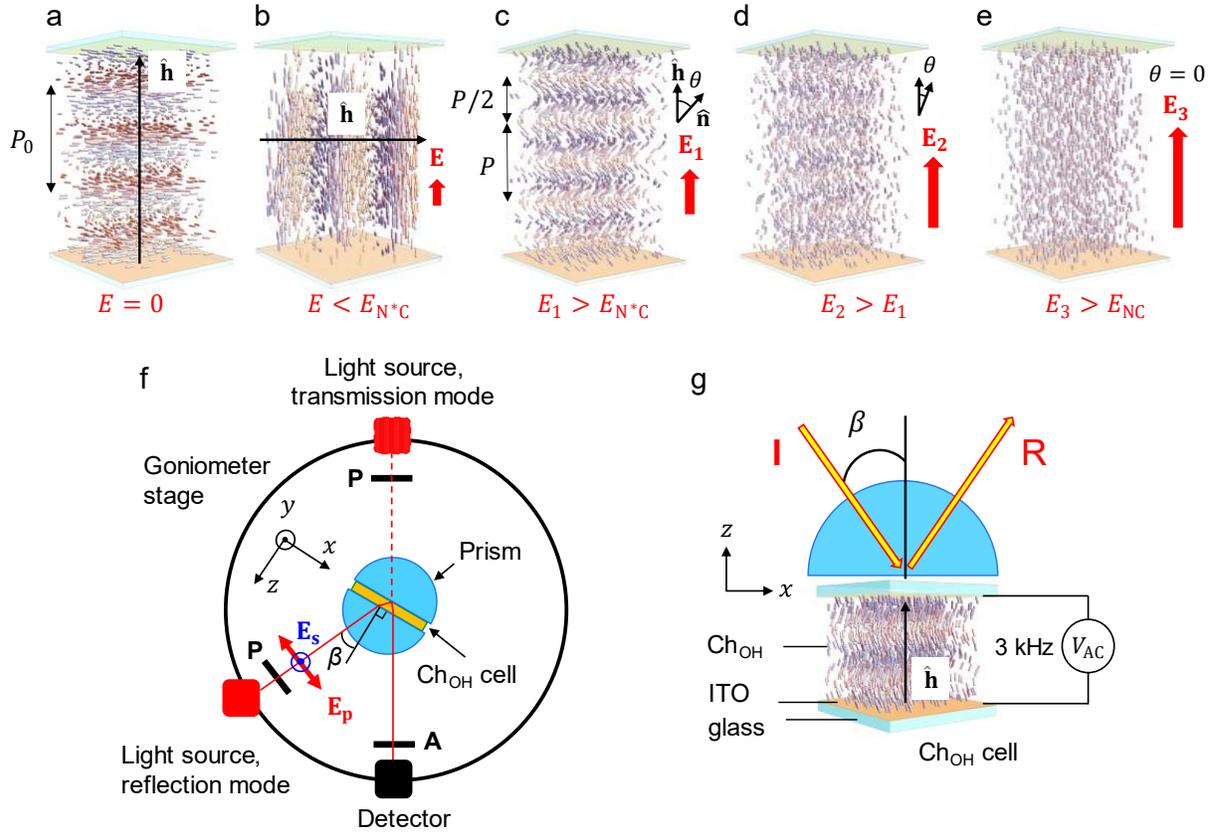

**Figure 1**. Schematic drawing of (a) a ground Ch state in the absence of the electric field, and (b)-(e) field-induced states in the $Ch_{OH}$ material: (b) a light-scattering "fingerprint" Ch state stabilized by a weak electric field $\mathbf{E} < E_{N^*C}$, (c),(d) $Ch_{OH}$ structure with a heliconical angle $\theta$ and a pitch $P$ both $E$-tunable in the range $E_{N^*C} \leq \mathbf{E} < E_{NC}$ (both $\theta$ and $P$ decrease when the field $\mathbf{E}$ increases), (e) an unwound state with $\theta = 0$ and the director $\hat{\mathbf{n}}$ aligned parallel to a strong applied field $\mathbf{E} \geq E_{NC}$. (f) Top view of the experimental setup. The $Ch_{OH}$ cell is mounted in the center of the goniometer stage with cell's $xy$ surface being perpendicular to the plane $xz$ of light incidence. The vectors $\mathbf{E}_p$ and $\mathbf{E}_s$ correspond to p- and s-polarizations of the incident light, respectively. The polarizers P control polarization of the incident beams, while the analyzer A is used to explore the polarization of reflected/transmitted beams. The angle of light incidence $\beta$ is measured from the normal $\hat{\mathbf{z}}$ to the $Ch_{OH}$ cell. (g) Schematic drawing of the $Ch_{OH}$ cell subject to an applied ac field, 3 kHz.

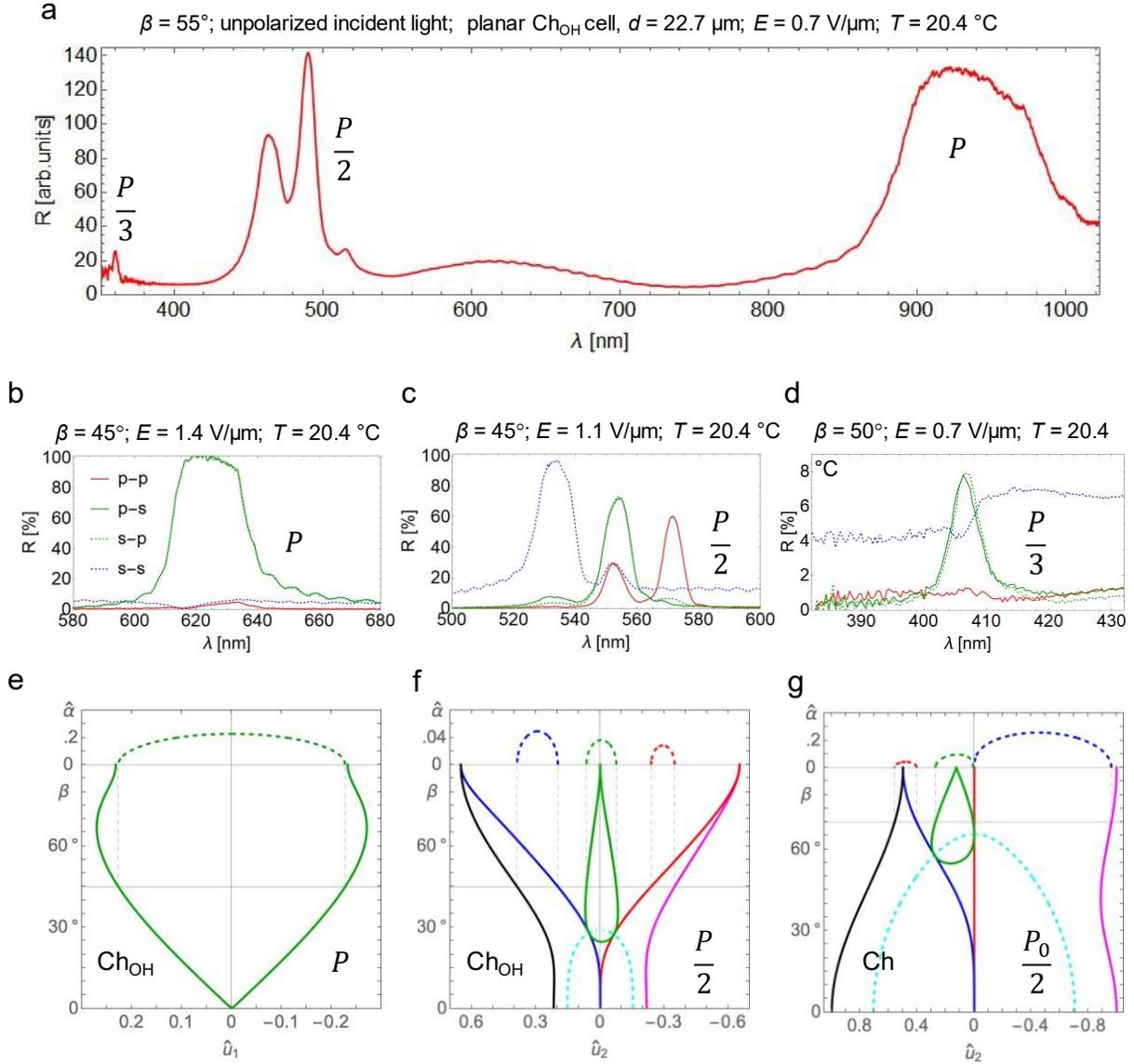

**Figure 2**. Multiple orders of Bragg reflection at Ch$_{OH}$. (a) Reflection from $P$, $P/2$, and $P/3$ observed simultaneously in the Vis and NIR spectral regions; unpolarized light; planar cell, $d = 22.7$ μm; angle of light incidence $\beta = 55°$; $E = 0.7$ V/μm. Polarization characteristics of reflections at (b) $P$, (c) $P/2$ and (d) $P/3$; spectra are presented for p-p, p-s, s-p, and s-s polarizations, where p-s and s-p states are overlapping; the first and second symbols indicate polarizations of the incident and reflected beams, respectively. The theoretically calculated dependencies of (e) the first-order and (f) the second-order Bragg bands of Ch$_{OH}$ on the incident angle $\beta$. The upper parts of the panels show $\hat{\alpha} = \operatorname{Im} k_{-,+}$ which determines the amplitudes of the Bragg bands for $\beta = 45°$, which corresponds to the experimental $\beta$ in (b) and (c). Dashed blue, green, and red lines correspond to s-s, s-p, and p-p polarizations, respectively; (g) The second order Bragg reflection band for a conventional Ch vs $\beta$; the upper parts of the panels show that the s-s and p-p bands switch sides as compared to the case of Ch$_{OH}$ in part (f).

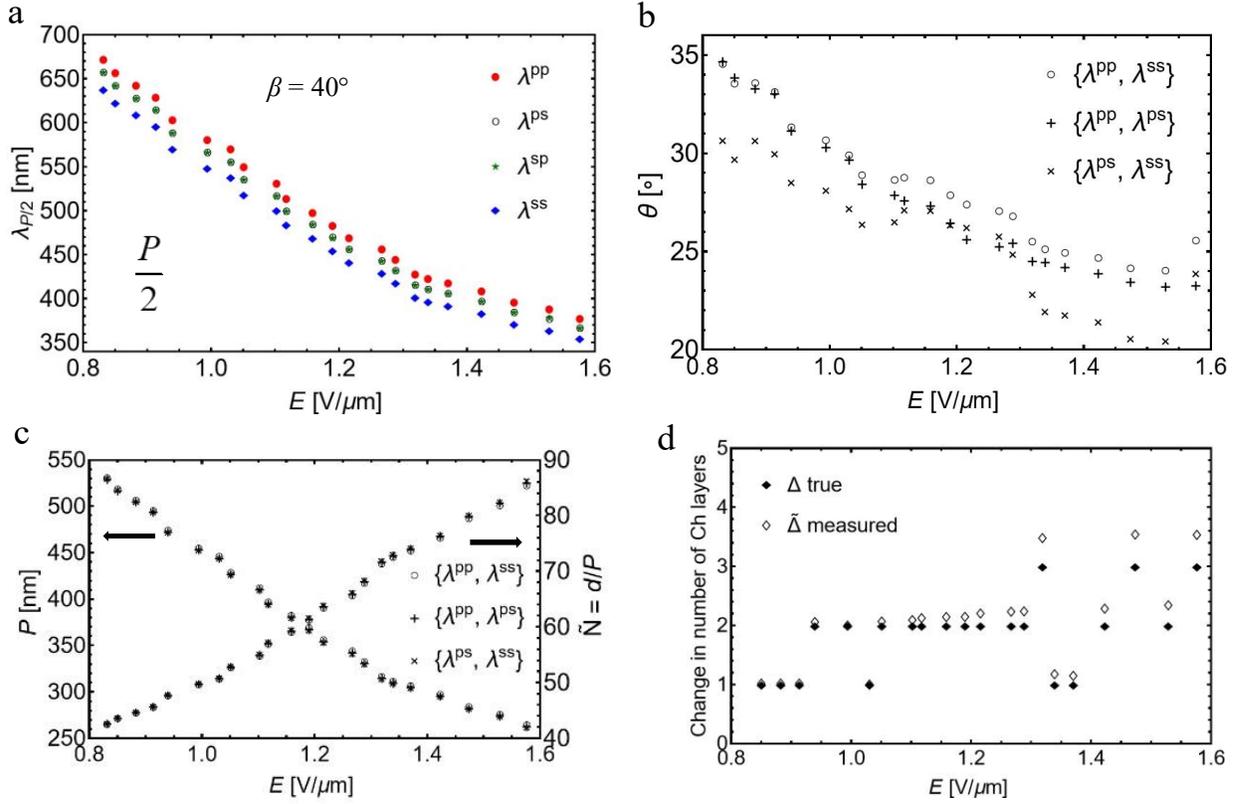

**Figure 3**. Electrically tunable (a) wavelength $\lambda_{P/2}$, (b) heliconical angle $\theta$, (c) pitch $P$ (left vertical axis) and the ratio $\tilde{N} = d/P$ (right vertical axis) measured in the planar cell, $d = 22.7$ μm, at an angle of incidence $\beta = 40°$. In (b) and (c), the values of $\theta$ and $P$ are determined from the Equation (32) solving for the pairs $\{\lambda^{pp}_{P/2}, \lambda^{ss}_{P/2}\}$, $\{\lambda^{pp}_{P/2}, \lambda^{ps}_{P/2}\}$, and $\{\lambda^{ps}_{P/2}, \lambda^{ss}_{P/2}\}$. (d) The change in number of Ch pseudolayers $\tilde{\Delta}$ is close to the integer at low fields, and deviates from the integer at stronger fields.

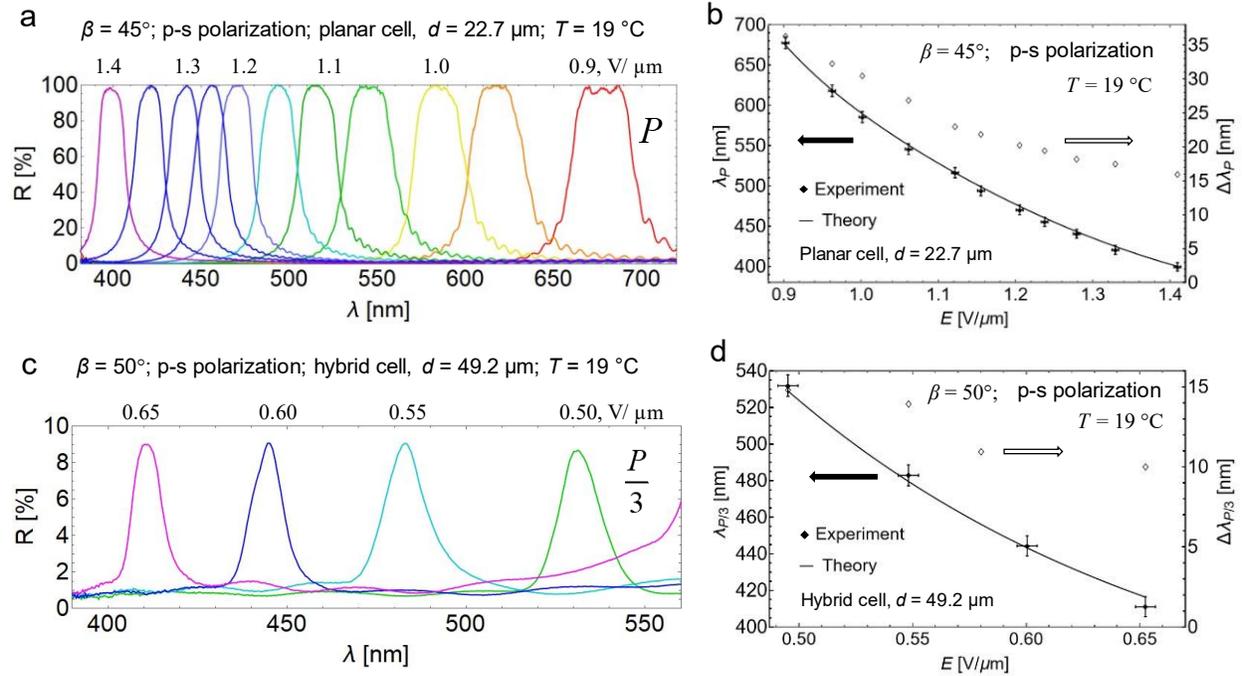

**Figure 4**. Electrically tunable reflection of light at (a) the Ch$_{OH}$ pitch $P$ and (c) one-third of the pitch $P/3$; RMS amplitudes of the field are indicated at the top of some bands; (b) Electric field dependency of the reflection maximum $\lambda_P$ and the bandwidth $\Delta\lambda_P$; planar cell, $d = 22.7$ μm; angle of light incidence $\beta = 45°$; (d) Similar $E$-field dependency of $\lambda_{P/3}$ and $\Delta\lambda_{P/3}$; hybrid aligned cell, $d = 49.2$ μm; $\beta = 50°$. In (b) and (d), the data are fitted to the theory using values of $\theta$ and $P$ determined from the Equation (32).

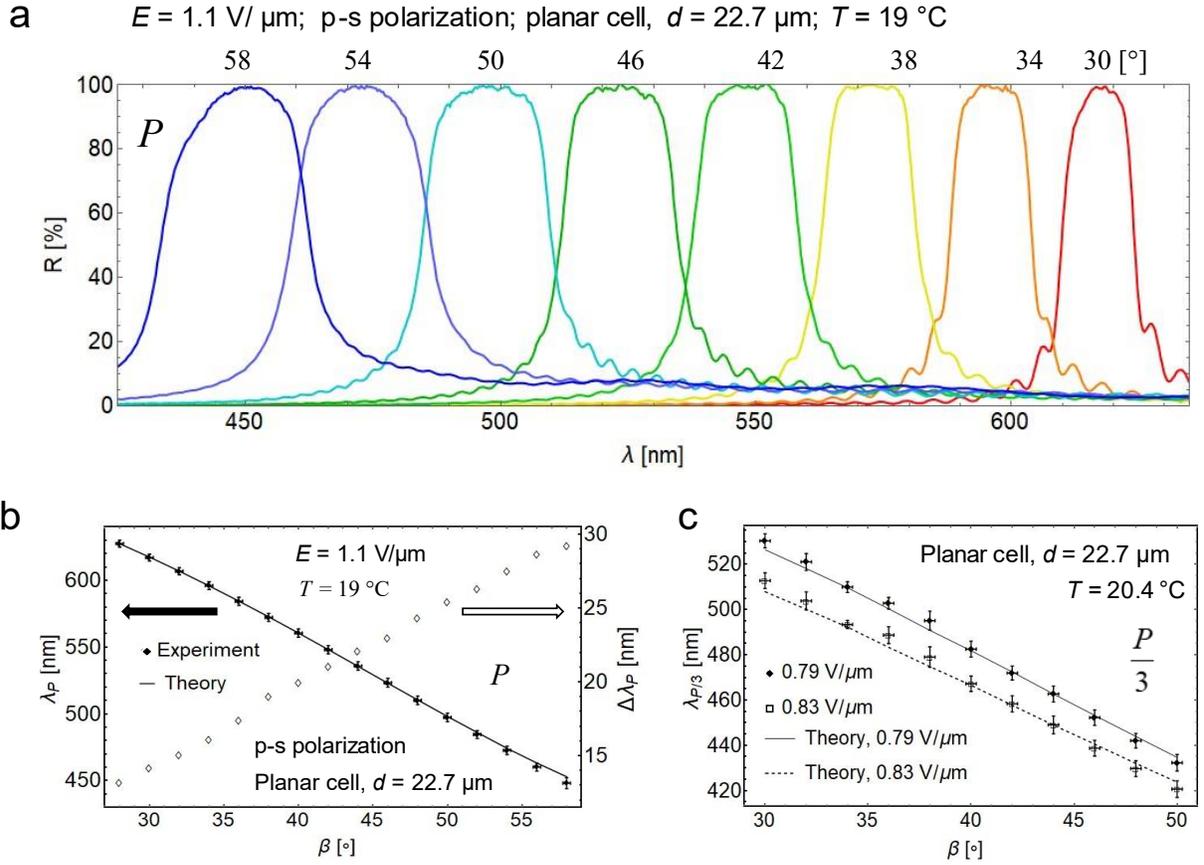

**Figure 5.** Control of peak wavelength by the angle of light incidence $\beta$: (a) Spectra of reflection at the full pitch $P$; the angle $\beta$ is shown above the Bragg peaks; $E = 1.1$ V/μm. (b) Wavelength $\lambda_P$ and bandwidth $\Delta\lambda_P$ of reflection at the full pitch $P$ vs. $\beta$; (c) Similarly, the reflection at $P/3$ vs. $\beta$ is shown for $E = 0.79$ V/μm (filled diamonds) and $E = 0.83$ V/μm (empty squares). In (b) and (c), the data are fitted to the theory using material parameters $P_0 = 1.7$ μm, $E_{NC} = 4.7$ V/μm, and $\kappa = 0.03$ measured previously [33].

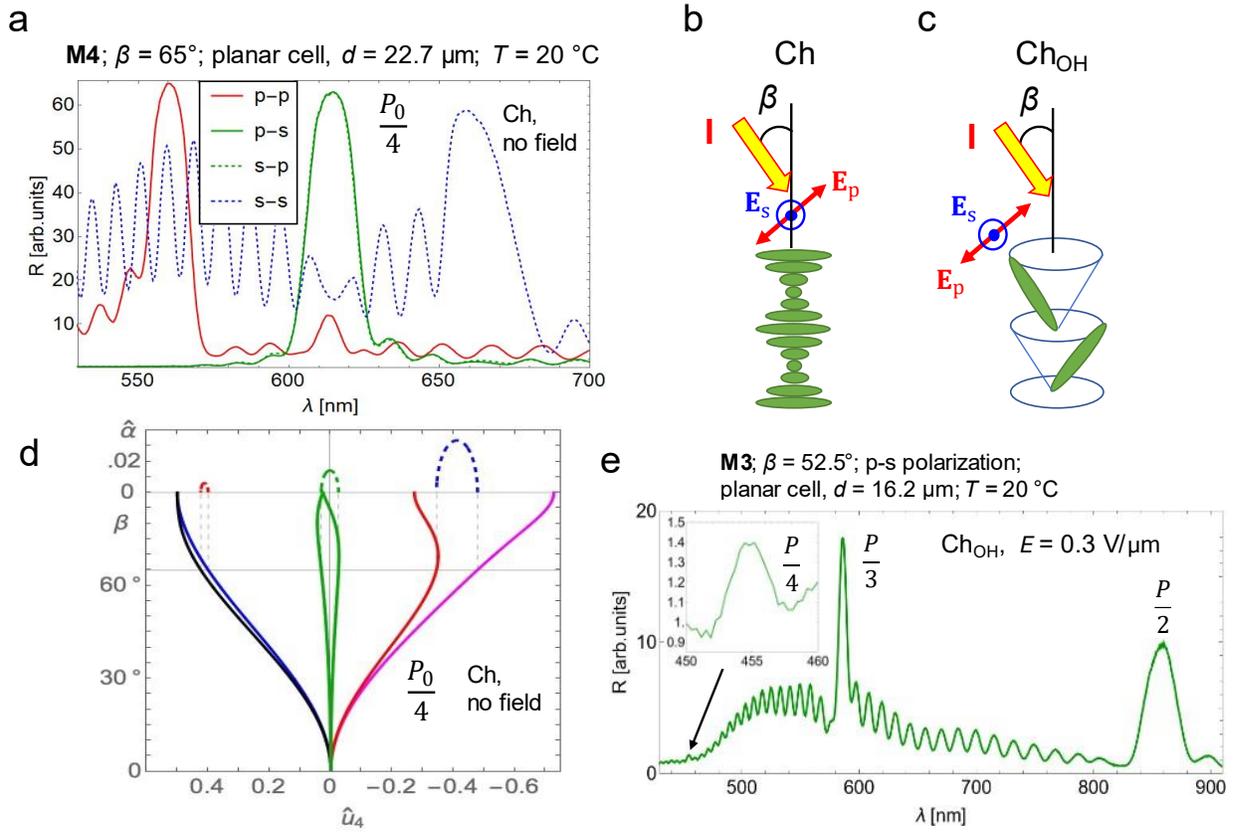

**Figure 6.** Bragg reflection at one quarter of a pitch in Ch and Ch$_{OH}$: (a) Reflection at $P_0/4$ in Ch measured in M4. The spectra are presented for p-p, p-s, s-p, and s-s polarizations, where p-s and s-p states are overlapping; the first and second symbols indicate polarizations of the incident and reflected beams, respectively. The swap in polarization characteristics of Ch$_{OH}$ and Ch are due to the difference between the effective refractive indices as illustrated in (b) and (c). Note the redshift of the p-p polarization away from the central peak in the Ch$_{OH}$, Fig.2c, and its blue shift in the Ch, Fig.6a. (d) The theoretically calculated dependency of the fourth-order Bragg band of a conventional Ch on the incident angle $\beta$. The upper parts of the panel show $\hat{\alpha} = \mathrm{Im}\, k_{-,+}$ which determines the amplitudes of the Bragg bands for $\beta = 65°$ (the same as the experimental $\beta$ in (a)). Dashed blue, green, and red lines correspond to s-s, s-p, and p-p polarizations, respectively; note that the theoretical bands are less symmetric and narrower than the experimental one; the features can be explained by an increased birefringence in the blue part of the spectrum and by director fluctuations ("noise"). (e) Bragg reflection of the fourth order $P/4$ in Ch$_{OH}$ measured in M3; p-s polarization; planar cell, $d = 16.2$ μm; $\beta = 52.5°$; $E = 0.3$ V/μm.

**Electrically tunable total reflection of light by oblique helicoidal cholesteric**

Olena S. Iadlovska[1,2], Kamal Thapa[1,2], Mojtaba Rajabi[1,2], Mateusz Mrukiewicz[1,4], Sergij V. Shiyanovskii[1,3], and Oleg D. Lavrentovich[1,2,3]*

[1]Advanced Materials and Liquid Crystal Institute, Kent State University, Kent, Ohio 44242, USA
[2]Department of Physics, Kent State University, Kent, Ohio 44242, USA
[3]Materials Science Graduate Program, Kent State University, Kent, Ohio 44242, USA
[4]Institute of Applied Physics, Military University of Technology, 00-908 Warsaw, Poland
*Author for correspondence: E-mail: olavrent@kent.edu, tel.: +1-330-672-4844.

**Refractive indices and birefringence**

Analysis of Bragg reflections requires a knowledge of the dispersion of refractive indices. In the N mixture, the ordinary $n_o$ and extraordinary $n_e$ refractive indices are measured at wavelengths 488, 532, and 633 nm using a wedge cell and the interference technique [1]. The wavelength selection is achieved using laser line color filters with 1 nm central bandwidth (Thorlabs, Inc.). The wedge cell is assembled from two ITO glass substrates coated with rubbed polyimide PI2555 (Nissan Chemicals, Ltd.) to achieve a planar alignment; the rubbing direction is perpendicular to the thickness gradient. The thickness of the thick end of the wedge cell is set by 100 μm spacers mixed with the adhesive NOA68, and there are no spacers at the glued wedge. The wedge angle is measured to be less than 1°. The cell is filled with the N mixture in isotropic phase, and the refractive indices are measured on cooling, Figure S1a.

The dispersion of refractive indices $n_e(\lambda)$ and $n_o(\lambda)$ is described by the Cauchy formula $n_{e,o}(\lambda) = A + B\lambda^{-2}$. Within the temperature range $19\,°C \leq T \leq 27\,°C$, both $n_e$ and $n_o$ are practically temperature-independent, with the mean values $n_e = 1.770$ at 488 nm, $n_e = 1.745$ at 532 nm, and $n_e = 1.717$ at 633 nm; $n_o = 1.556$ at 488 nm, $n_o = 1.545$ at 532 nm, and $n_o = 1.533$ at 633 nm. The polynomial coefficients found by fitting the $n_e(\lambda)$ dispersion are $A = 1.64$, $B = 3.08 \cdot 10^{-2}$ μm². Similarly, for $n_o(\lambda)$, $A = 1.50$, $B = 1.33 \cdot 10^{-2}$ μm², Figure S1a.

To verify whether the measured N data match the properties of the Ch mixture, we determined Ch birefringence through the measurements of optical retardance $\Gamma$ of a thin ($d = 5.3$ μm) planar cell with the director unwound by a strong in-plane ac (3 kHz) electric field $E > E_{NC}$, applied to two ITO stripe electrodes separated by a 100 μm gap, Figure S1b. $\Gamma$ is measured at 475, 535, and 655 nm using the Exicor MicroImager System (Hinds Instruments). The birefringence of the unwound Ch is then calculated as $\Delta n = \Gamma/d$, Figure S1b. The $\Delta n$ values of the Ch and N phases are in a very good agreement with each other, Figure S1c, differing by 0.005 or less. In what follows we use the dispersion data $n_e(\lambda)$ and $n_o(\lambda)$ obtained for the N mixture.

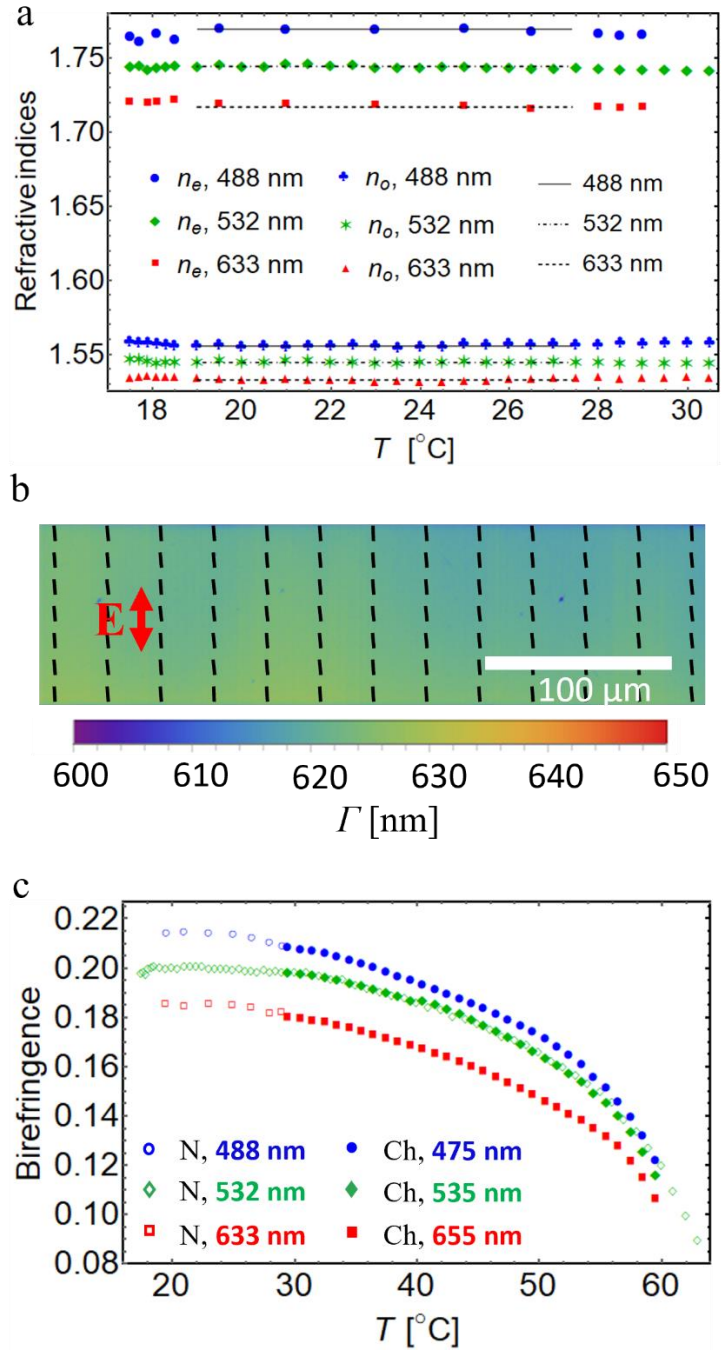

**Figure S1.** (a) Temperature dependencies of the refractive indices measured in N mixture. (b) MicroImager map of the optical retardance of the Ch$_{OH}$ cell subject to a high in-plane electric field $E > E_{NC}$. (c) Temperature dependence of the birefringence measured in N (empty symbols) and Ch (filled symbols). The size of all data symbols exceeds the error of measurements.

**Off-field Ch pitch $P_0$**

The intrinsic Ch pitch $P_0$ is measured at $T = 20$ °C using two approaches: (i) by analyzing the separation between dislocations of the Burgers vector $P_0/2$ in the Grandjean-Cano wedge [2] and (ii) by Bragg reflection of an obliquely incident light at the planar Ch structure, Figure S2. The measured values of $P_0$ are $1.69 \pm 0.02$ μm and $1.67 \pm 0.03$ μm, respectively. In what follows, we use $P_0 = 1.7$ μm at $T = 20$ °C.

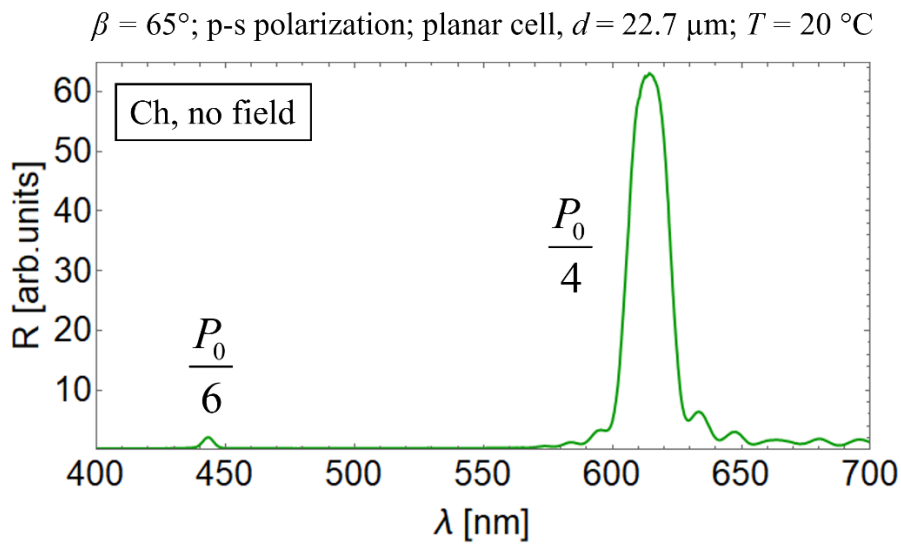

**Figure S2.** Bragg reflections at $P_0/4$ and $P_0/6$ by a planar cell, $d = 16$ μm, used to determine the Ch pitch $P_0$; oblique incidence of light; p-s polarization; $T = 20$ °C.